\documentclass{llncs}
\pdfoutput=1
\usepackage{epsfig, endnotes, url}
\usepackage{color, cite, graphicx}
\usepackage{mathtools, amsmath, amsfonts}
\usepackage{diagbox, booktabs, colortbl}
\usepackage{multirow, caption, subcaption}
\usepackage{breakurl, framed}

{
\makeatletter
\newcommand{\@chapapp}{\relax}%
\makeatother

\usepackage[title]{appendix}
\captionsetup[subfigure]{justification=raggedright}

\newcommand{\ignore}[1]{}

\usepackage{array}
\newcolumntype{P}[1]{>{\centering\arraybackslash}p{#1}}

\hyphenation{op-tical net-works semi-conduc-tor}

\begin{document}
\title{\Large \bf ABC: A Cryptocurrency-Focused Threat Modeling Framework}

\author{Ghada Almashaqbeh\inst{1}\thanks{Supported by NSF CCF-1423306.} \and
Allison Bishop\inst{1,2}\thanks{Supported by NSF CCF-1423306 and NSF CNS-1552932.} \and
Justin Cappos\inst{3}}
\authorrunning{G. Almashaqbeh et al.}
\institute{Columbia University, NY, USA
\email{\{ghada, allison\}@cs.columbia.edu}\\ 
\and 
Proof Trading, NY, USA\\
\and
New York University, NY, USA
\email{jcappos@nyu.edu}}

\maketitle

% Use the following at camera-ready time to suppress page numbers.
% Comment it out when you first submit the paper for review.
%\thispagestyle{empty}

\begin{abstract}
Cryptocurrencies are an emerging economic force, but there are 
concerns about their security. This is due, in part, to complex collusion
cases and new threat vectors that could be missed 
by conventional security assessment strategies. To address  
these issues, we propose ABC, an {\bf A}sset-{\bf B}ased {\bf C}ryptocurrency-focused 
threat modeling framework capable of identifying such risks. 
ABC's key innovation is the use of collusion matrices.  A collusion matrix
forces a threat model to cover a large space of threat cases
while simultaneously manages this process to prevent it from being overly
complex. Moreover, ABC derives a system-specific threat 
categories that account for the financial aspects and the new asset types 
that cryptocurrencies introduce. We demonstrate that ABC is effective 
by conducting a user study and by presenting real-world use cases. 
The user study showed that around 71$\%$ of those 
who used ABC were able to identify financial security threats, as compared 
to only 13$\%$ of participants who used the popular framework STRIDE. 
The use cases further attest to the usefulness of ABC's tools for both 
cryptocurrency-based systems, as well as a cloud native security technology. 
This shows the potential of ABC as an effective security assessment technique for 
various types of large-scale distributed systems.
\end{abstract}

\section{Introduction} 
\label{intro} 
Cryptocurrencies and blockchain technologies are an emerging 
economic force. Since Bitcoin emerged in 2009~\cite{bitcoin}, the 
number of 
these ``digital currencies" has grown into the thousands in 2019, with 
a total market capital exceeding $\$$380 billion~\cite{cap-market}. 
As the value of these currencies has grown, their goals also changing. 
Early systems focused only on providing 
a virtual currency exchange medium~\cite{bitcoin, ripple}, but nowadays there 
is increasing interest in providing other types of distributed services, 
such as computation outsourcing~\cite{golem} or file 
storage~\cite{filecoin}, on top of this medium. These newer 
applications suggest increased adoption of 
cryptocurrency-based systems in the coming years.

Yet, despite the many advantages they offer ---   
decentralization, transparency, and lowered service costs --- there is still 
a big gap between the promise of cryptocurrencies and their performance in 
practice. A major stumbling block is the perception that these systems are 
not secure, and the large number of 
security breaches announced in the past few years give  
credence to these doubts~\cite{dao, eth-hack, ico-hack-b, exchange-hack, 
exchange-hack-b, bitfloor-hack, enigma-hack, poloniex-hack, 
bitstamp-hack, steemit-hack, veritaseum-hack, stellar-hack, japan-exchange-hack, 
mine-hack, benebit-scam, bitcoingold-double-spending, 
binance-hack, korea-exchange-hack, bitcoin-cash-hack}. Therefore, a better understanding  
of the security of cryptocurrencies is needed in order  
to ensure their safe deployment in emerging applications, and their 
continued adoption.

The best practice for designing a secure system requires a
threat modeling step to investigate potential 
security risks. Such a model can guide developers in deploying the 
proper countermeasures during the design 
phase, and in assessing the system's security level in the after-design stage. 
Although threat modeling has been thoroughly studied 
in the literature, existing paradigms primarily target  
software applications~\cite{Howard06} or distributed systems that have 
a small number of participant types~\cite{Myagmar05}. These threat 
modeling techniques were not designed
to be scalable for a set of diverse, mutually distrustful parties as is found 
in cryptocurrencies. Such  
systems, especially those providing distributed services, 
consist of parties that play different roles (miners, clients, and 
different types of servers), and an attacker may control any subset of  
these parties.  Adding to the complexity, the attacker may seek to target 
any role in the system and may launch a diverse array of attacks with different
intended outcomes.  As these sets grow, the complexity of reasoning about and managing threat cases becomes unwieldy.

To address these issues, we propose ABC, an {\bf A}sset-{\bf B}ased 
{\bf C}ryptocurrency-focused threat modeling framework. ABC introduces 
a novel technique, called a collusion matrix, that allows users to investigate 
the full threat space and manage 
its complexity. A collusion matrix is a comprehensive investigation and threat 
enumeration tool that directly addresses collusion by accounting for all possible 
sets of attacker and target parties. ABC reduces  
the combinatorial growth of these cases by ruling out irrelevant scenarios   
and merging threat cases that have the same effect. This explicit consideration of 
attacker collusion is particularly important in permissionless 
cryptocurrencies that allow anyone to join.

ABC's models are also tailored to better consider the threat domain of 
cryptocurrencies. 
This is done by introducing new threat categories that account for the   
financial motivations of attackers and new asset types, i.e., critical  
components, these systems introduce.
ABC identifies these categories by listing the assets in a system, 
such as the blockchain and the peer-to-peer network, and outlining what 
entails secure behavior for each one. Then, the threat categories are defined as any violation  
of the security requirements for these assets. This approach produces a series of 
system-specific 
threat classes as opposed to an a priori-fixed list of generalized  
ones. Another feature of ABC is acknowledging that financial incentives and 
economic analysis can play major roles in other steps in the design process. 
These tools can be used in risk assessment and in mitigating some types 
of attacks that cannot be neutralized cryptographically.

To demonstrate the framework's effectiveness, we  
conducted a user study and prepared use cases. The study  
compared the performance of subjects in building threat models using  
ABC and the popular framework STRIDE. Among 
the obtained results, we found that around 71$\%$ of 
those who applied ABC were able to identify financial 
threats in a cryptocurrency system, as compared to less than 
13$\%$ of those applying STRIDE. In addition, while none of the 
STRIDE session participants spotted collusion between attackers, 
around 46$\%$ of those who used ABC identified these scenarios.

For the use cases, we applied ABC to four real-world 
systems: Bitcoin~\cite{bitcoin}, Filecoin~\cite{filecoin}, CacheCash, 
and SPIFFE~\cite{spiffe}. These cases 
attest to the usefulness of ABC's tools, as they 
integrated well into CacheCash's design phase, and revealed  
several missing threat scenarios in the public design of Filecoin. Furthermore, 
ABC proved to be advantageous in a collaboration between one of this paper's 
authors and a team working on assessing and improving the 
security of SPIFFE, a cloud-based identity production framework 
hosted by the Linux Foundation. Using ABC, the SPIFFE group was able to 
reason about all threat cases in a 
systematized way, determine the critical threats, and then 
prioritize mitigation actions accordingly. This   
confirms the potential of ABC as an effective tool for 
assessing and improving security not only for cryptocurrency-based systems, but 
for large-scale distributed systems in general. 

\section{Related Work} 
\label{related-work} 
To orient readers to the current state-of-the-art in threat modeling, 
we summarize here some prior work done in this 
area. We also highlight relevant works  
on threat identification and security analysis in cryptocurrencies. \\

\noindent{\bf Threat modeling frameworks.} The STRIDE framework, 
developed by Microsoft as part of its Security Development Lifecycle 
(SDL), is one of the earliest and most popular works in this field 
\cite{Howard06, Torr05}. STRIDE is an acronym of the threat categories 
the framework covers, namely, Spoofing, Tampering, Repudiation, 
Information disclosure, Denial of service, and Elevation of privilege.  
This framework is a multistep procedure that involves understanding the 
software application functionality, capturing its operation flow using data 
flow diagrams (DFDs), mapping the components of these DFDs to the  
threat categories mentioned previously, and employing threat tree patterns 
to derive concrete threat cases.

Though several solutions have extended STRIDE to accommodate 
more complex systems~\cite{Myagmar05}, and cover other security 
requirements, e.g., privacy~\cite{Deng11, Luna12}, its model does not fit 
cryptocurrencies. Another study~\cite{Vandervort15}, in which the authors 
extended STRIDE's threat categories to handle    
Bitcoin-like community currencies, bears out 
this premise. However, their approach 
targets only community fund operation; more 
modifications would be needed to handle other 
components of the currency exchange medium, and 
other types of distributed services.

Other paradigms have pursued slightly different approaches. KAOS 
\cite{Van04} is a goal-oriented requirements engineering framework 
that has been extended to cover security. It analyzes the anti-security goals 
of a system to identify the types of threats that represent potential compromises. T-MAP~\cite{Chen07} is a value-driven framework that targets 
commercial off-the-shelf systems. It identifies all attack paths and assigns  
them severity weights based on the organization (or business) 
policy to help in evaluating security practices. ANOA~\cite{Backes13} is a 
generic framework to define and analyze anonymity of   
communication networks, while the frameworks presented in~\cite{Andel08, Hollick17} 
target the secure design of data routing protocols. Finally, other works 
build specialized threat models 
for specific classes of distributed systems, e.g., storage systems~\cite{Hasan05}, 
virtual directory services~\cite{Claycomb09}, and 
unmanned aerial vehicle systems~\cite{Javaid12}, rather 
than introducing a framework.

The aforementioned works indicate  
that different types of systems have different requirements when 
performing threat modeling. This reinforces the idea that emerging 
systems, such as cryptocurrencies, 
need specialized threat modeling tools. \\

\noindent{\bf Security analysis of cryptocurrencies.} Most of the work 
done so far in this category can be divided into two classes. The first class 
formalizes the security properties of consensus protocols and  
blockchains~\cite{Pass17, Garay15}, with the goal of providing a security 
notion and a rigorous framework 
to prove security of blockchain-based systems, 
whereas the second class discusses specific 
security attacks on cryptocurrencies. For example, in a series of studies 
on Bitcoin, Bonneau et al.~\cite{Bonneau15} present several security threats, 
Androulaki et al.~\cite{Androulaki13} evaluate its anonymity property, 
Gervais et al.~\cite{Gervais15} study how tampering with the 
network links affects participants' view of 
the blockchain, and Kroll et al.~\cite{Kroll13} study the economics of 
Bitcoin mining and the effect of miner financial incentives on its security.

Other works dealing with different types of cryptocurrencies include Luu et al. 
\cite{Luu15, Luu16} who focus on 
security threats to smart contracts in Ethereum~\cite{ethereum},    
Kosba et al.~\cite{Kosba16} who provide a model for decentralized smart 
contracts that preserve user's privacy, and  
Sanchez et al.~\cite{Sanchez16, Sanchez17} who analyze 
the security of Ripple~\cite{ripple} and the linkability of wallets 
and transactions in its network. Moreover, a recent empirical study,    
by Moser et al.~\cite{Moser18}, shows that 
the transactions in the privacy preserving cryptocurrency Monero~\cite{monero} 
are traceable and that their real inputs can be identified. And another, by 
Kappos et al.~\cite{Kappos18}, shows that 
the anonymity set of the private transactions in Zcash~\cite{zcash} can be shrunk 
using a simple heuristic derived from coin usage patterns.

While these attack descriptions are very useful, they only 
outline specific threat scenarios for a given system. Our goal, however, is to 
develop a framework that allows reasoning about the full set of potential 
attacks facing any cryptocurrency-based system.

\section{Stepping through the ABC Framework} 
\label{abc} 
Having highlighted the need for a cryptocurrency-specific threat 
modeling framework, we now present the ABC model that answers to this 
need. As a systematized approach, applying ABC starts by   
understanding the functionality of the cryptocurrency system under 
design with a focus on its asset types and 
the financial motivations of the participants (Section~\ref{step1}). This 
is followed by identifying the impactful threat categories and 
mapping them to the system 
assets (Section~\ref{step2}). After that, ABC directs system designers 
to extract concrete attack scenarios using a new tool called  
a collusion matrix, which helps in exploring and analyzing the full 
threat space (Section~\ref{step3}). Lastly, ABC 
acknowledges that financial incentives 
affect other design steps, including risk 
assessment and threat mitigation (Section~\ref{step4}).

To make the discussion easier to follow, we illustrate the ABC 
process by describing its application to the following simplified system, 
which was inspired by Golem~\cite{golem}: \\

{\bf CompuCoin} is a cryptocurrency that provides a distributed 
computation outsourcing service. Parties with excessive CPU 
power may join the system as servers to perform computations 
on demand for others. Clients submit computation jobs to  
servers, wait for the results and proofs of correctness, and 
then pay these servers with cryptocurrency tokens. The mining process 
in CompuCoin is tied to the amount of service 
provided to the system. That is, the probability of a server being selected   
to mine the next block on the blockchain is proportional to the 
amount of computation it has performed during a specific period. \\

The full threat model for CompuCoin is available online~\cite{material}.  
Several excerpts from this model are embedded in the discussion of ABC 
steps that follows.

\subsection{System Model Characterization} 
\label{step1}
Understanding the system is an essential step in the 
threat modeling process. A misleading or incomplete system description   
can lead a designer to overlook serious threats and/or incorporate irrelevant ones. 
Therefore, an accurate system model must outline the use 
scenarios of the system, the 
assumptions on which it relies, and any dependencies on external 
services. In addition, the model must be aware of all 
participant roles, and any possible motivation each  
might have to attack the system. For the latter, evaluators need to consider 
how the financial interests of these participants shape their behaviors.

Moreover, a system model must define the critical components 
that need to be protected from attackers. These components 
represent the assets that would compromise the whole system if 
attacked. To capture the features of the system, ABC 
identifies these assets 
based on functionality. In detail, ABC divides the 
system into modules, and labels the valuable components of each 
module as assets, which could be concrete or abstract 
resources~\cite{Myagmar05}. For example, the blockchain and 
the currency can be considered 
concrete assets, while preserving user privacy would be    
an abstract asset.

Finally, a system model includes graphic illustrations of its work 
flow. For distributed systems, it is useful to draw network 
models~\cite{Myagmar05} in which system  
modules are represented by graphs showing all    
participants and assets, and the 
interactions between them. These graphs  
are helpful when enumerating the concrete threat scenarios as we will  
see in Section~\ref{step3}.

\begin{figure}[t!]
\centerline{
\includegraphics[height= 1.6in, width = 0.9\columnwidth]{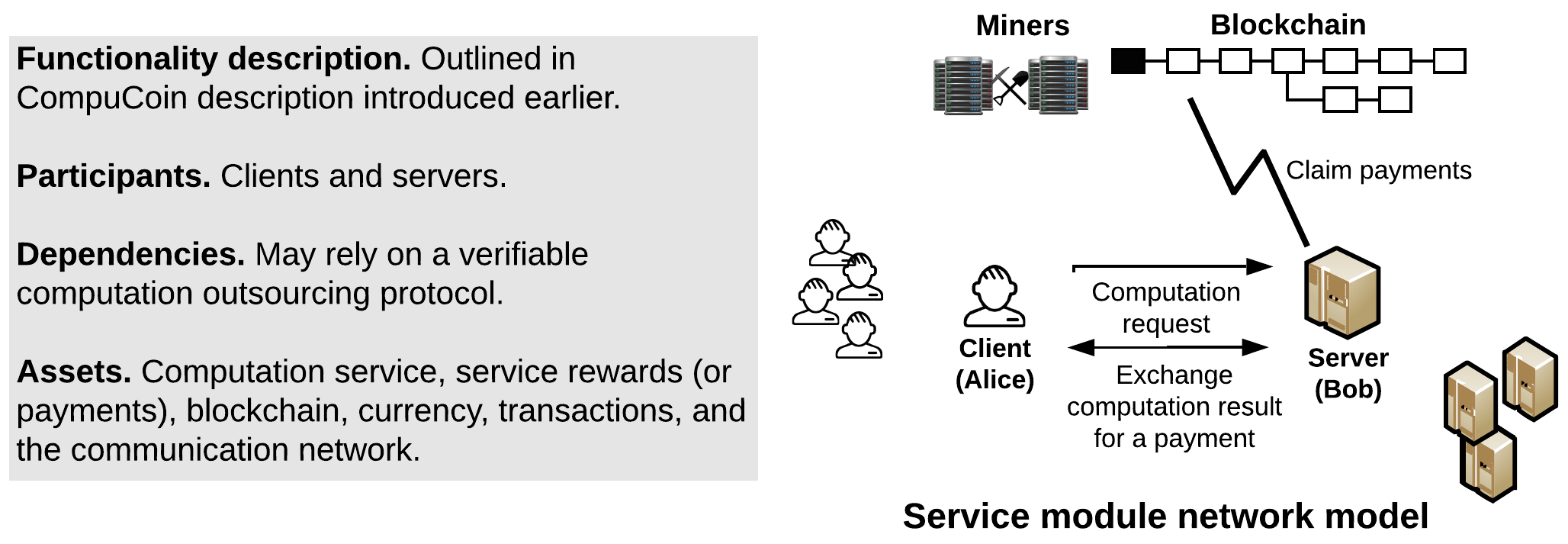}}
\caption{System model characterization of CompuCoin. }
\label{system-model}
\end{figure}

{\bf Running example application.} Figure~\ref{system-model} illustrates 
how this step would look in CompuCoin. It shows the 
various components of the system model, in addition to a network model 
of the computation outsourcing service. To cover the full functionality of the 
system, other network models would be needed to capture components, such as 
the mining and consensus processes. Furthermore, the asset list in this figure  
is not exhaustive, and is limited by the rather brief 
description provided for CompuCoin.

As shown in the figure, anyone can join CompuCoin as 
clients and servers, with servers also filling the role of miners. Dependency 
on other systems may include reliance on a 
verifiable outsourcing computation protocol, e.g. \cite{Parno13}. 
 In terms of assets, one may 
define three in CompuCoin: the \emph{Service} promised to clients, the 
\emph{Payments} used to compensate for the service, and the 
\emph{Currency Exchange Medium} that covers four sub-assets 
(in light of the extended review of Bitcoin \cite{Bonneau15}): 
the blockchain, currency, transactions, and the 
communication network that connects the parties together. Here one may 
merge the currency with the transactions in one asset, as transactions are 
usually the currency tokens that state the coin's ownership. Another option 
is to merge currency with the payment asset to cover all currency flow in 
the system. However, we believe that a fine-grained division provides 
a more comprehensive treatment when identifying threats.

\subsection{Threat Category Identification} 
\label{step2}
After understanding the system model, the next step 
is to identify the broad threat categories that must be investigated. 
For each system component, system analysts outline all threat classes that 
may apply. Here ABC steps away from the conventional practice of using an 
a priori-fixed list, and instead uses an adaptive approach inspired by  
requirements engineering~\cite{Deng11}. This approach defines threats as violations 
of system security goals. Given that assets are the 
target of security breaches, ABC defines these threat classes as violations  
of asset security requirements. 
This allows deriving system-specific threat 
categories because ABC identifies the assets in a way that aligns  
with the functionality of the system under design.

Accordingly, in this step, an evaluator examines each asset 
and applies the following procedure to identify its threat classes:
\begin{itemize}
\itemsep0em
\item Define what constitutes secure behavior for the asset, and 
use that knowledge to derive its security requirements. 
These requirements include all conditions that, if met, would 
render the asset secure. For example, CompuCoin's servers provide 
a computation outsourcing service and collect payments in return. One may 
consider the service payment asset secure if: a) servers are 
rewarded properly for their work, and b) that they earned 
the payments they collected.

\item Define the threat categories of an asset as violations of its 
security requirements. Tying this to the above example, the service 
payment asset would have the following threat classes: 
service slacking, where a server collects payments without performing 
all the promised work, and service theft, where a client obtains service 
for a lower payment than the agreed upon amount. 
\end{itemize}

The previous steps are highly dependent on how 
system analysts define the security 
properties of an asset, especially if there is no agreed-upon definition 
in the literature. For example, several works  
studied the security of the 
blockchain and the consensus protocol~\cite{Bonneau15, Garay15, Pass17}. 
Yet, there is no unified 
security notion for the service asset because each type 
may have different requirements.

\begin{table}[t!]
\caption{CompuCoin threat categories.} 
\scriptsize{
\label{threat-cat}
\centering
\setlength\extrarowheight{2pt}
\begin{tabular}{|p{0.2\columnwidth}  | p{0.8\columnwidth} |}\hline\hline
{\bf Asset} & {\bf Security Threat Category} \\ [0.5ex] \hline\hline

\multirow{4}{*}{Service}  &  Service corruption (provide corrupted service for clients). \\ [0.5ex] \cline{2-2}
&  Denial of service (make the service unavailable to legitimate users). \\ [0.5ex] \cline{2-2}
&  Information disclosure (service content/related data are public). \\ [0.5ex] \cline{2-2}
                                       & Repudiation (the server can deny a service it delivered).  \\[0.5ex] \hline
                                       
\multirow{2}{*}{Service}  &  Service slacking (a server collects payments without performing all the promised work). \\ [0.5ex] \cline{2-2}
     \vspace{-10pt}payments                                  & Service theft (a client obtains correct service for a lower payment than the agreed upon amount).  \\[0.5ex] \hline

\multirow{3}{*}{Blockchain}  & Inconsistency (honest miners hold copies of the blockchain that may differ  beyond the unconfirmed blocks).  \\ [0.5ex] \cline{2-2}
 & Invalid block adoption (the blockchain contains invalid blocks that do not follow the system specifications).  \\ [0.5ex] \cline{2-2}
                                         &  Biased mining (a miner pretends to expend the needed resources for mining to be elected to extend the blockchain). \\ [0.5ex]  \hline
                                         
\multirow{3}{*}{Transactions}  & Repudiation (an attacker denies issuing transactions).  \\ [0.5ex] \cline{2-2}
       & Tampering (an attacker manipulates the transactions in the system).  \\ [0.5ex] \cline{2-2}
                                         &  Deanonymization (an attacker exploits transaction linkability and violates users' anonymity). \\ [0.5ex]  \hline

{Currency}  & Currency theft (an attacker steals currency from others in the system). \\ [0.5ex]  \hline
{Network}  & Denial of service (interrupt the operation of the underlying network). \\ [0.5ex]  \hline
\end{tabular}}
\end{table}

{\bf Running example application.} Applying this step to  
CompuCoin produced the threat categories listed in 
Table~\ref{threat-cat} (the detailed process of deriving the 
security requirements and negating them to produce the listed 
threats for each of these assets is presented  
in Appendix~\ref{threat-categories}). We found 
this table useful when building threat models for  
all the use cases introduced in Section~\ref{experiences}, where 
we mapped the listed categories to the assets in each system. In 
this mapping process, we found that some threat types were 
not applicable due to 
the absence of some assets. Notably, Bitcoin's only assets are the 
ones related to the currency exchange medium. On the other hand, 
other systems required replicating some of these categories 
among all instances of an asset, e.g., in Filecoin all service asset threats  
were replicated for the two service types this system provides, namely, file 
storage and retrieval. This shows how the system characteristics affect 
the threat category identification step in ABC.

\subsection{Threat Scenario Enumeration and Reduction} 
\label{step3}
Once the threat categories have been identified, the next step
is to enumerate concrete attack scenarios under each threat 
type. It is important in this step to be as comprehensive as possible by 
considering all potential attackers and target 
parties, as well as the set of actions attackers may follow, and the capabilities they 
must posses, to achieve their 
goals. This also involves considering collusion 
between several participants who may cooperate 
to attack the system.

Detecting collusion is particularly important in cryptocurrencies. The presence of 
monetary incentives may motivate attackers to 
collude in more ways than traditional distributed systems. The 
popular centralization problem caused by mining pools attests to this 
fact, as when these pools collude they can perform devastating attacks. 
Even miners may collude by accepting, or rejecting, updates 
on the network protocol which leads to hard forks in the system. 
ABC can detect these and other collusion cases at early stages of 
the system design.

To achieve this, ABC introduces collusion matrices that instruct     
analysts to enumerate all collusion cases, and reason   
about the feasibility of all threat scenarios in the system. 
A collusion matrix is two-dimensional, 
with the rows representing potential attackers and the columns 
representing the target parties. For the rows 
we list all participant roles in the system, both individually and in 
every possible combination. We also add a category called  ``external" 
that represents all entities outside the system. The same is done for the 
columns, with the exception that ``external" is excluded. By definition, an 
external party is not part of the system, and hence, can not be a target. 
Each cell in these matrices represents a potential threat case to be 
investigated.

\begin{figure}[t!]
\centerline{
\includegraphics[height= 1.8in, width = 1.0\columnwidth]{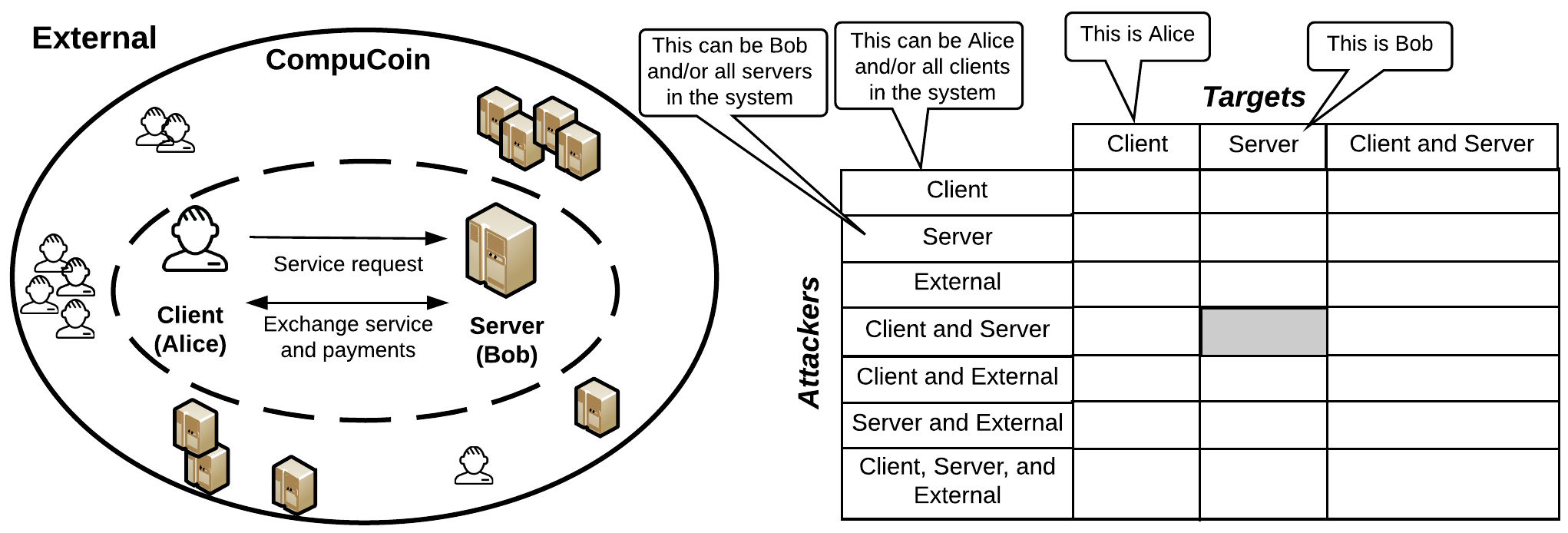}}
\caption{Collusion matrix of service theft threat in CompuCoin.} 
\label{collusion-matrix-example}
\end{figure}

An example of a collusion matrix for the service theft threat in 
CompuCoin is shown in 
Figure~\ref{collusion-matrix-example}. The dashed ellipse in the 
accompanying network model encloses a service session, which is 
an interaction between a server and a client. 
Any entry with multiple parties on the attacker  
side in this matrix indicates collusion. Note that a participant 
label may represent slightly different roles depending on where it is 
placed. For example, in Figure~\ref{collusion-matrix-example} 
the label ``server" on the target side corresponds 
to a single server (i.e., Bob) since a service session involves only one 
server. However, the label ``server" on the attacker side represents all 
servers in the system, including Bob. Hence, the cell in grey shade in 
Figure~\ref{collusion-matrix-example} does not suggest that a server 
colludes to attack itself, but instead, it represents the case where   
other servers collude with Alice against Bob.

For each threat category mapped to the assets in the system, a separate 
collusion matrix is created and analyzed as follows: 

\noindent{\bf 1) Enumeration:} In this step, system analysts  
examine each cell and enumerate all strategies that attackers with 
specific capabilities 
can use against the target parties, and documenting the process. It is useful to 
consider the network 
model of the system components as they show the interactions between 
the participants and the system assets. \\

\vspace{-6pt}
\noindent{\bf 2) Reduction:} While examining each cell, system analysts 
reduce the number of threat cases by:
\begin{itemize}
\vspace{-3pt}
\item Eliminating cells representing scenarios that will not produce a 
threat to the system. This consists of crossing out the eliminated cells
and documenting the rationale for elimination.  For example, in 
Figure~\ref{collusion-matrix-example}, the cells that have the client as 
a target are irrelevant to the service theft threat. This is because a client 
does not provide a service to others. Other 
cases can also be crossed 
out if they are neutralized by system assumptions or by early design 
choices. For example, requiring all transactions to 
be signed by their originators rules out transaction repudiation and tampering 
attacks.

\item Merging together scenarios (and the corresponding cells) that have the 
same effect, or those  
that do not become stronger with collusion. For example, 
in Figure~\ref{collusion-matrix-example}, the grey shaded cell in which Alice is 
colluding with other servers to avoid paying Bob is reduced to the case that 
Alice is a sole attacker. This is because only Alice pays for the service she receives 
from Bob, while other servers are not part of the protocol\footnote{The case that these clients 
drop/withhold these payments in collusion with Alice is part of 
other threats, such as DoS attack.}.
\end{itemize}

\noindent{\bf 3) Documentation:} System analysts should 
document all threat scenarios that remain after the reduction step. 
That is, each documented case should outline the attack description, 
the target parties and assets, the attacker(s), the flow of actions, 
all preconditions that make the attack feasible, and the 
reasons behind merges and deletions (if any). \\

\vspace{-6pt}
The overall number of matrices 
and the size of each matrix depend on the system parameters, 
such as number of participant roles and assets. 
The above reduction step eliminates a substantial number of cells in a 
documented, principled way, saving time and effort.

\begin{figure}[t!]
\centerline{
\includegraphics[height= 1.8in, width = 0.8\columnwidth]{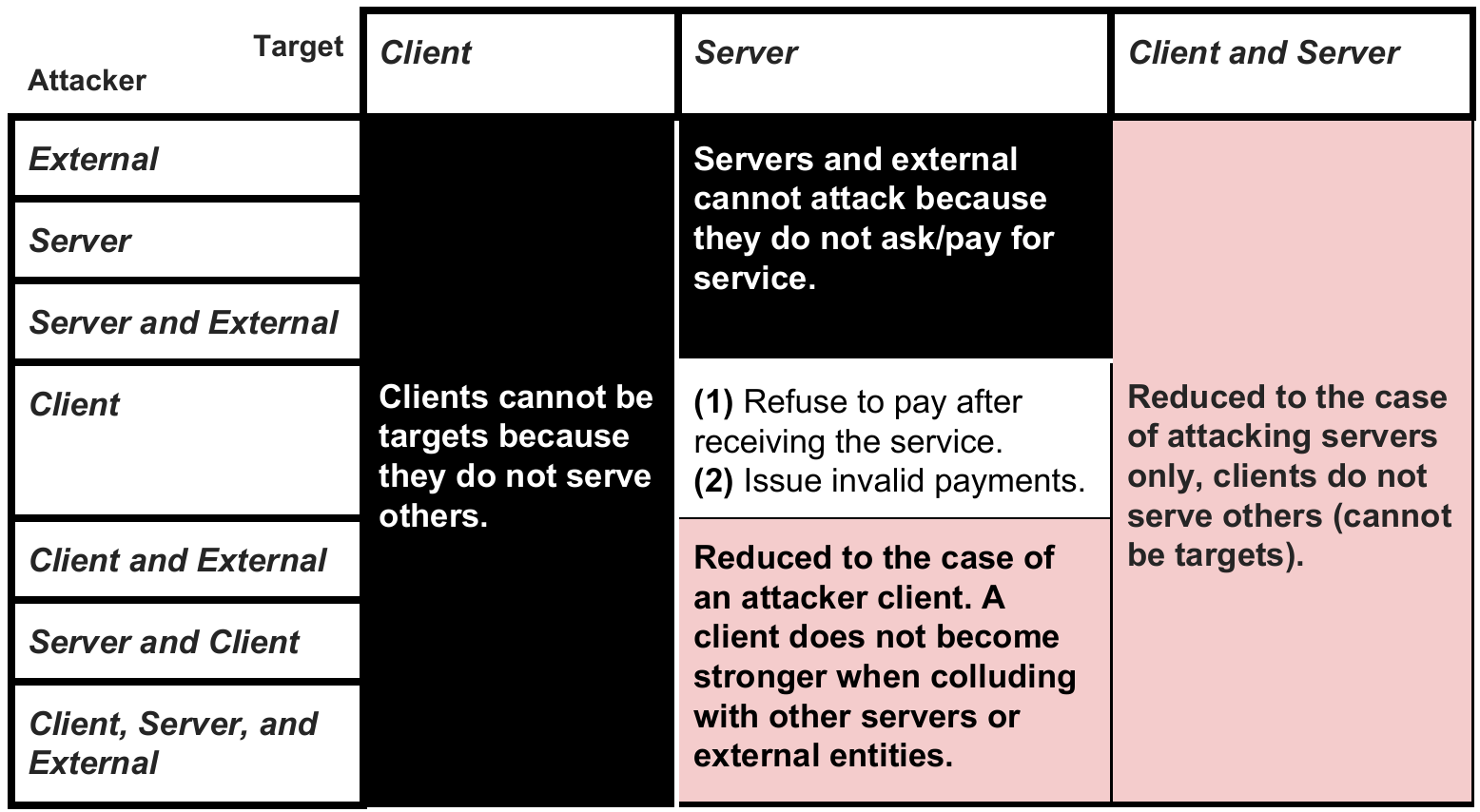}}
\caption{Analyzing the collusion matrix of the service theft threat in CompuCoin.}
\label{service-theft-matrix-compucoin}
\end{figure}

{\bf Running example application.} The CompuCoin threat model has 11  
collusion matrices~\cite{material}. We 
present one of them here: the service 
theft threat collusion matrix as illustrated in Figure~\ref{service-theft-matrix-compucoin}. 
As shown, 21 cells can be reduced to just 2 threat scenarios (merged 
and ruled-out cells are displayed 
in pink and black shades, respectively). In this matrix, ten cases have 
been ruled out. These include all cells under the column with the ``client" header, 
for the reasons explained previously, and the first three cells under 
the column with the ``server" 
header. This is because ``external"  
and/or ``server" cannot be attackers because they do not ask/pay for 
the service\footnote{One may 
say that an external may join the system as a client to perform the attack. This 
case is covered under the client role in the matrix.}.

Ten other merged cases are shown in 
Figure~\ref{service-theft-matrix-compucoin}. This includes all cells under the column 
with the ``client and server" header, which are reduced to attacking only 
servers. This is again because clients do not serve others. The rest of the 
merges cover the last three cells in the column 
with the ``Server" header. In these cells, a client is colluding with an external 
entity and/or other servers to make the target server lose payments.  
Such collusion will not make a client stronger  as all these parties can do is drop/withhold 
payments, an option already covered under DoS threat. Hence, all 
these cells are reduced  
to the case of a solo client attacker.

\subsection{Risk Assessment and Threat Mitigation} 
\label{step4}
The outcome of the threat modeling process, i.e., the documented list of 
impactful threat cases, gives the designers  
a guiding map to secure the system. During 
this process, it is useful to prioritize threats based on the amount of 
damage they can cause and the likelihood that an attacker has   
the required capabilities to carry out them. This falls under the 
purview of risk management, a separate task from threat modeling, 
carried out using frameworks like  
DREAD~\cite{Howard06} or OCTAVE~\cite{Alberts02}.

ABC integrates with risk management by leveraging existing techniques for
threat mitigation. For example, many threat
vectors can be addressed using rational financial 
incentives that are often called detect-and-punish mechanisms. 
That is, when a cheating incident is detected, the miners  
punish the attacker financially. Others can rely on designing algorithms 
that when executed in a malicious way consume more resources, i.e., 
incur a larger cost, than 
an honest execution. These approaches can use a game 
theoretic approach~\cite{Tadelis13} to set the design parameters in a way that makes 
cheating unprofitable. By modeling interactions between 
the players as an economic game, the financial gain of all player   
strategies can be computed. Then,  
the parameters are configured to make  
honest behaviors more profitable than cheating.

The same procedure can be used 
to quantify the damage these financial threats may cause. In other words, 
a threat that could give the attacker a big payoff should be prioritized 
over a threat that yields minimal profits. This reinforces the idea that cryptocurrencies 
require an expanded model for exploring risks and countering them.

{\bf Running example application.} To illustrate this step  
in CompuCoin, we consider the distilled threat scenarios found in 
Figure~\ref{service-theft-matrix-compucoin}. Both threats can be 
neutralized financially by designing proper techniques to make  
the client lock the payments in an escrow and create a penalty deposit before 
asking for any service. The client 
loses the penalty deposit if it  
should cheat, perhaps by issuing invalid payments that carry its  
signature. The deposit amount needs to be at least equal to the 
maximum additional payoff a cheating client may obtain as compared to 
an honest client. This makes cheating 
unprofitable, and hence, unappealing to rational clients.

\section{Evaluation} 
\label{evaluation} 
To evaluate the effectiveness of ABC, we set up an empirical experiment 
that compares how it performs against STRIDE~\cite{Howard06},  
a widely used threat modeling framework. We chose STRIDE for this 
comparison because 
it is a popular example of the type of a model a system designer will turn to 
in the absence of a cryptocurrency-specific 
framework~\cite{Vandervort15}. The experiment took the form of a 
user study in which participants were asked to build threat 
models for a simple cryptocurrency system using one of these two 
frameworks. Our primary goal was to test whether financial incentives 
and collusion could influence the type of threats discovered. Thus, our 
evaluation focuses on answering the following questions: 
\begin{enumerate} 
\itemsep0em
\item Does a threat modeling framework affect how subjects 
characterize a system model? 

\item Do the threat categories of each framework influence  
the broad threat classes identified by the subjects?

\item Do participants build more accurate threat models when using 
ABC than when using STRIDE? 

\item Do participants find the ABC/STRIDE method easy to use in 
completing the study?
\end{enumerate}

In what follows, we discuss the study methodology and some 
of the insights drawn from the findings.

\subsection{Methodology} 
We recruited 53 participants, primarily masters students in systems security 
programs. We used five subjects as a pilot group to 
test and refine our materials. The remaining 48 participants were divided 
randomly into two 
groups of 24, one of which built the threat model with STRIDE, whereas  
the other used ABC.

Each testing session spanned three hours and was divided into 
two parts: a group tutorial and individual 
completion of threat models. The group tutorial started with a 20 minute 
overview of cryptocurrencies, followed by a one-hour 
training in the framework to apply. The ABC tutorial contained a summary 
of the steps found in this 
paper, and for STRIDE, we prepared a tutorial based on material found 
in \cite{Torr05, Howard06, Deng11, user-guide16}. The 
participants were then given a 25 minute break to 
reduce any fatigue effects. The session resumed with a 15 minute overview of  
ArchiveCoin, the system for which subjects will  
build a threat model. ArchiveCoin is a simplified 
Filecoin~\cite{filecoin}-inspired cryptocurrency system that focuses mainly on the 
service and its rewards in order to fit the
study session period.

During the 
remaining hour of the study session, individuals worked independently 
to complete a threat model. Given that the allocated time 
was short, we asked the subjects to look into just one 
threat category in Step 3, namely, the service 
theft of file retrieval. This category was not used in the clarifying 
examples of the tutorials to avoid biasing the results. Participants 
performed Steps 1 and 2 (system model characterization and threat 
category identification), and then submitted 
their answers. Only at this point were they given the materials for 
Step 3, in which they were asked to elicit threat scenarios for  
service theft of file retrieval. This was done so that 
participants who missed this threat when answering Step 2  
could not alter their responses. At the end, the participants were asked 
to fill out a short questionnaire 
in which they rated how easy or hard it was to apply the threat modeling 
framework they employed. Our study instrument and all supporting 
materials are available online~\cite{material}.

\subsection{Findings} 
In this section, we present the study results using the four questions 
outlined previously as a guideline.

\subsubsection{Effect on System Model Characterization}
In the first step of each threat modeling framework, the subjects 
were asked to characterize the system model by 
defining its modules, its assets, and the participant roles, in addition to 
drawing either a network model of the system, in case of ABC, or a 
data flow diagram (DFD), in case of STRIDE.
To quantify the influence of the framework on this step, we compute 
the subject scores using   
reference threat models we built for ArchiveCoin prior to the study\footnote{We built two 
reference models, one 
using STRIDE and one using ABC to evaluate the responses of each 
framework session. Nonetheless, both models produced the same list of 
elicited threat cases in the last step.}. We 
report these scores after normalization, meaning that we 
divide them by the maximum score value one may obtain 
when answering everything correctly.

\begin{figure}[t!]
\centerline{
\includegraphics[height= 2.2in, width =0.8\columnwidth]{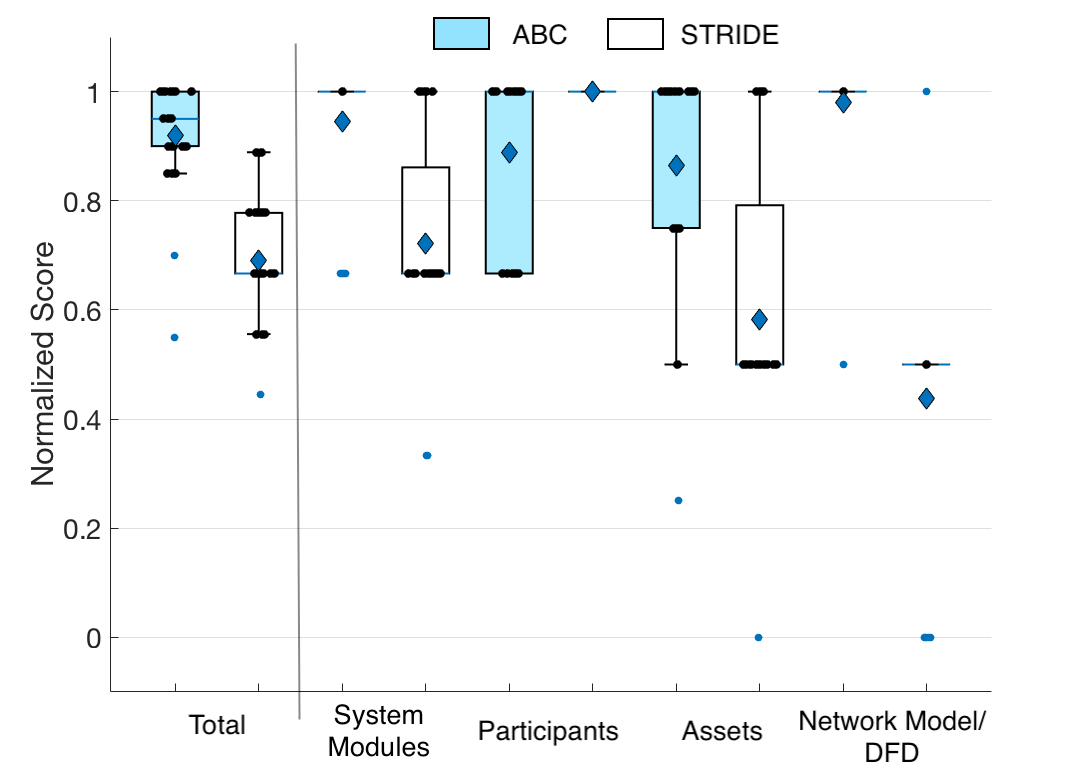}}
\caption{Subject scores for Step 1. Diamonds indicate the mean.}
\label{step1-scores}
\end{figure}

\begin{figure}[t!]
\centerline{
\includegraphics[height= 1.8in, width =0.6\columnwidth]{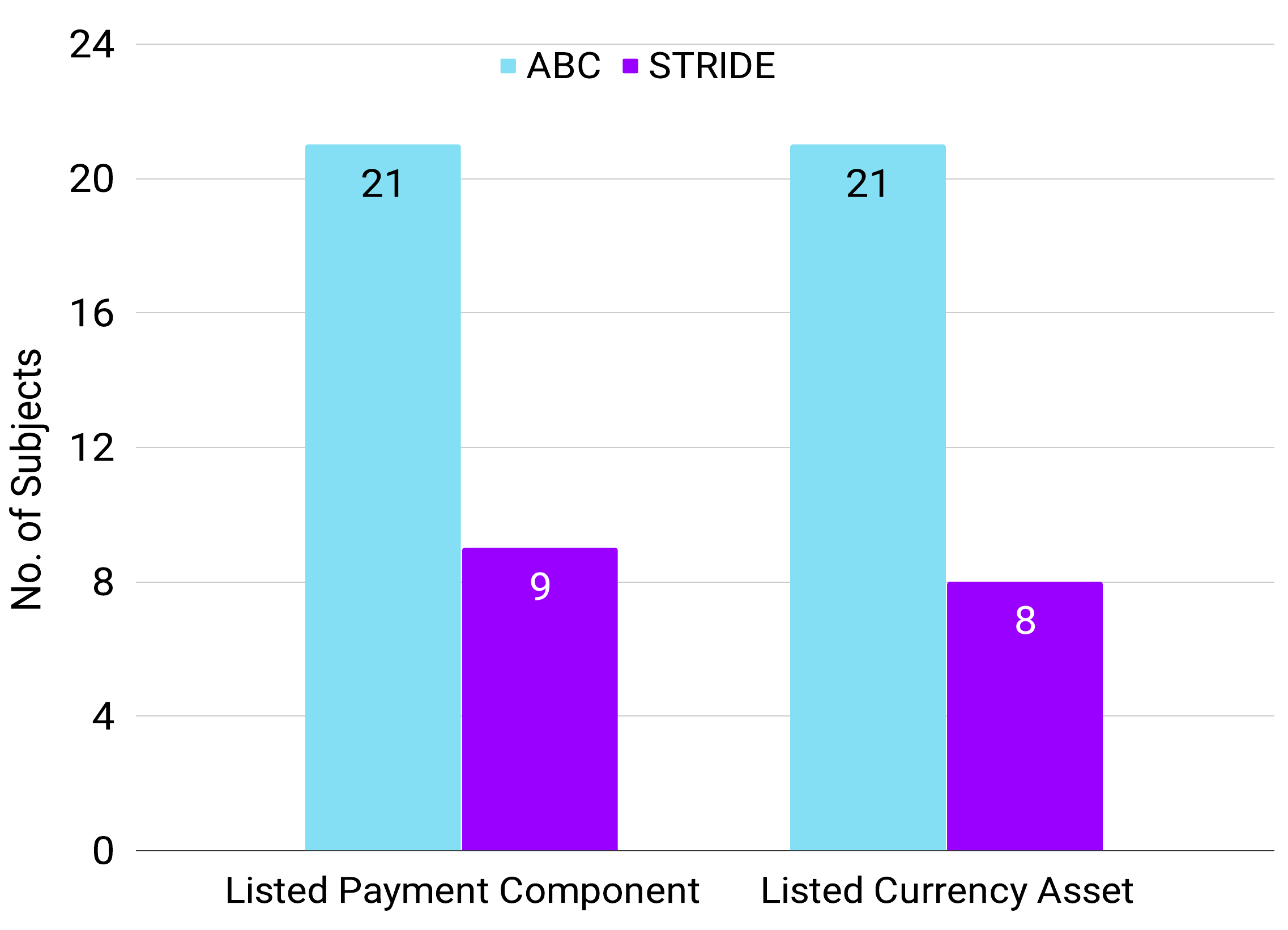}}
\caption{Subject frequency of identifying payment related modules and assets.}
\label{payment-module-asset}
\end{figure}

The results for Step 1 are found in Figure~\ref{step1-scores}\footnote{This 
figure is a box plot \cite{boxplot}, 
which displays the distribution of the data points by showing the 
maximum and minimum values (the whiskers 
above and below the box), the median (horizontal line inside the box), 
and the data points that span the first to third quartiles (the box itself).  
In case 
most of these points are very close this box is suppressed into a 
line.}. As shown, ABC scored higher than STRIDE, with total 
average values of 0.92 and 0.69, respectively. Analyzing the responses 
for the sub-steps in Step 1 revealed several interesting observations. The 
first one is related to identifying the financial assets and modules in the 
system (depicted in Figure~\ref{payment-module-asset}). As the figure 
shows, several subjects who 
applied STRIDE did not identify the payments (or currency) as an asset. Instead, 
their focus was on the user files stored in the system. Similarly, most of them 
did not identify the payment process as a system module, and focused 
only on file storage and retrieval processes. On the other hand, most of the subjects in 
the ABC session identified these financial related assets and modules. These 
results indicate that employing 
conventional threat modeling frameworks, instead of ones that are 
customized for monetary-incentivized systems, could  
lead evaluators to neglect the financial aspects of the system. This, in 
turn, could cause important threat cases to be overlooked, and thus, 
leave the system vulnerable to attacks.

The second observation is related to how subjects defined the participant roles in 
the system. As shown in Figure~\ref{step1-scores}, STRIDE achieved an 
average score of 1 in this category as compared to  
0.89 for ABC. All STRIDE session subjects defined the 
participant roles correctly, which in that model, included only clients 
and servers. For ABC, 
although its tutorial mentioned that an ``external'' entity 
must be considered among the participant roles, not all the subjects in that session  
recorded this role in their responses. This points to an 
important observation. Evaluators may only consider the insider 
attackers that interact with the system, and forget that external 
entities could be also motivated to, and capable of, an attack. Considering 
external attackers affects not only what concrete threat scenarios 
are elicited in Step 3, but also what threat categories are identified in Step 2. 
Therefore, more emphasis  
needs to be placed on this role early on in the threat modeling process.

Lastly, the third observation is related to how the framework influenced  
the way subjects represented the system modules graphically. 
Figure~\ref{step1-scores} shows that the average 
scores for the network model/DFD sub-step were 
found to be 0.98 and 0.44 for ABC and STRIDE, respectively. STRIDE 
session subjects struggled to draw a DFD for ArchiveCoin because such 
a representation is more suitable for software applications than distributed  
systems. On the other hand, ABC's use of network models made  
this task easier for its session subjects, and almost all of them sketched 
diagrams correctly. As mentioned previously, this graphic representation 
helps in eliciting the concrete threat scenarios in the system (Step 3), and 
hence, inaccurate diagrams may affect the outcome of this process.

\subsubsection{Effect on Threat Category Identification}
In Step 2 of both frameworks, the subjects were
asked to define the broad threat categories to be investigated. As 
part of the study material, participants who applied 
STRIDE were given its threat category list, along with the 
component mapping table found in the STRIDE user 
guide~\cite{user-guide16}. Similarly, participants who applied ABC were given the 
list found in Table~\ref{threat-cat} (covering only the service and 
service reward assets). The subjects in both groups defined the 
categories to be considered for ArchiveCoin by  
mapping these lists either to the system assets (in case of ABC), or to 
the DFD components (in case of STRIDE). The reference 
models we built indicated that the mapping outcome for both frameworks would 
include the following threat classes: service corruption, DoS, information 
disclosure, service slacking and theft for both service types that 
ArchiveCoin provides (file storage and retrieval).

\begin{figure}[t!]
\centerline{
\includegraphics[height= 1.8in, width =0.4\columnwidth]{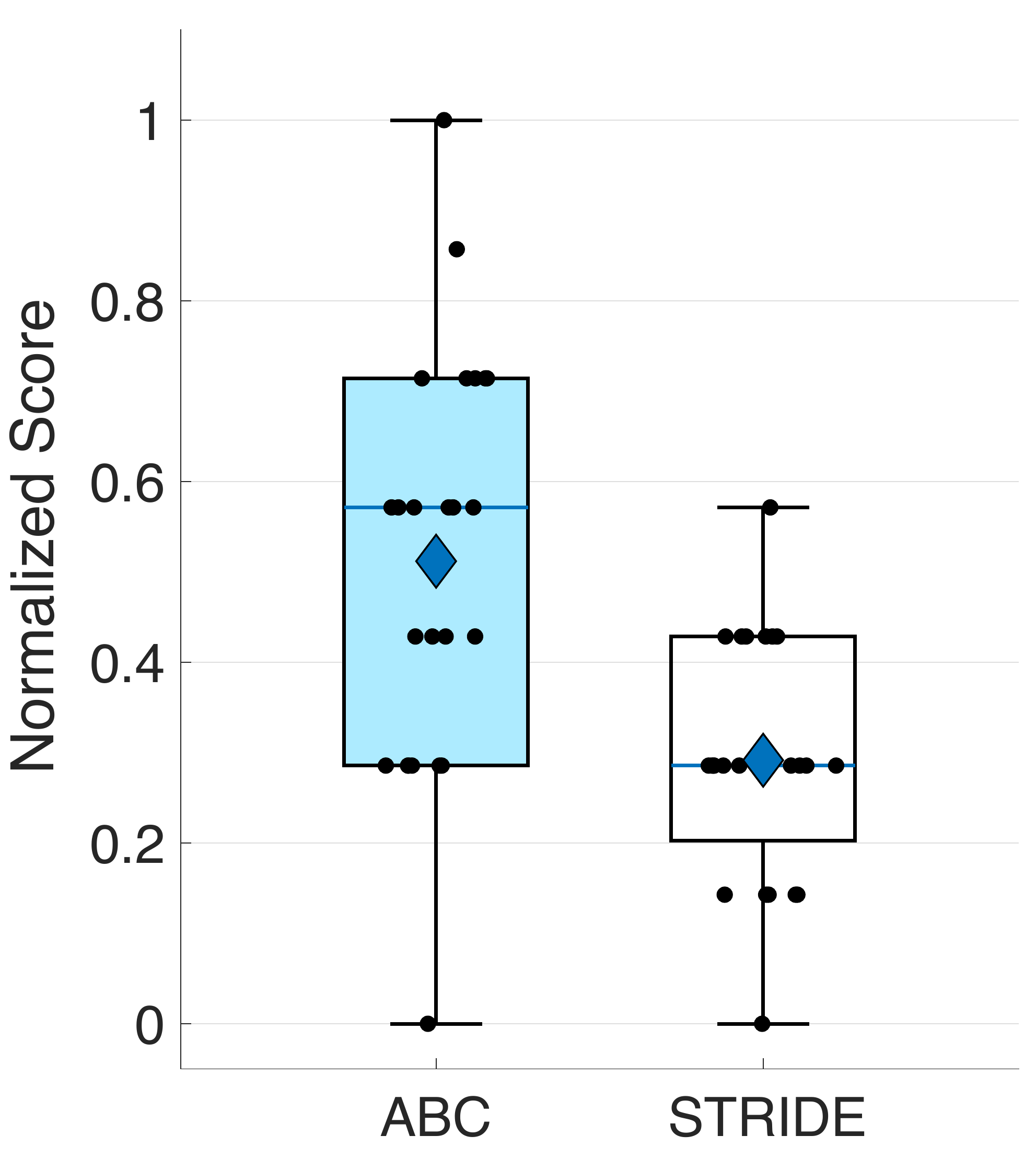}}
\caption{ABC and STRIDE scores for Step 2. Diamonds indicate the mean.}
\label{step2-scores}
\end{figure}

Based on the scores for the threat identification step found in 
Figure~\ref{step2-scores}, the cryptocurrency-tailored 
categories of ABC made it easier for the study participants  
to identify the threat categories in question as compared to STRIDE. 
This is despite the subjects having little 
experience with cryptocurrency-based systems. The average score for ABC 
subjects is around 0.51, compared to 0.29 for STRIDE (note these scores 
are normalized as mentioned before). The generalized categories 
used by STRIDE fit software applications well, but they do not suit 
monetary-incentivized 
distributed systems. System analysts, using these generalized categories,  
would need to expend more time and effort 
in order to identify the more specific threat classes of interest.

\begin{figure}[t!]
\centerline{
\includegraphics[height= 2.1in, width =0.8\columnwidth]{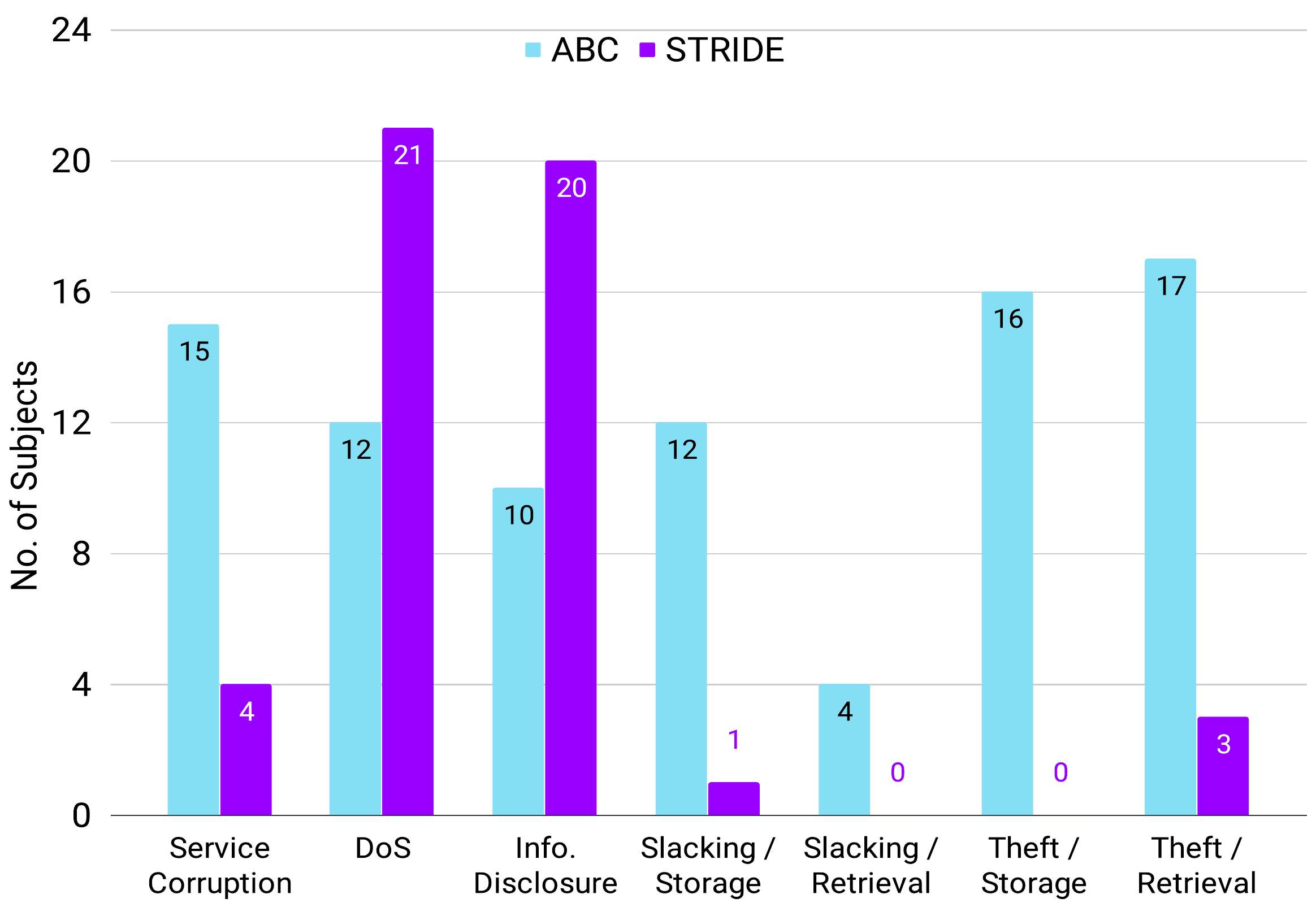}}
\caption{Subject frequency of threat category identification.}
\label{step2-freq}
\end{figure}

To provide more insights about this step, we analyzed the number of subjects who 
identified each threat category that need to be investigated. The results are 
depicted in Figure~\ref{step2-freq}.
We found that STRIDE's subjects   
are ahead of ABC's session participants for both DoS and information 
disclosure threats. As shown in Figure~\ref{step2-freq}, around 88\% 
and 83\% of STRIDE subjects 
identified these categories, respectively, while around 50\% and 42\% 
of ABC's subjects did so. Although we do not have a precise justification for 
this outcome, we think that this can be attributed to the threat 
category table of STRIDE, which 
thoroughly explains these categories and provides detailed attack examples. 
Hence, we believe that the ABC tutorial needs to stress 
these threats and explain them in greater depth.

In contrast, ABC is ahead of STRIDE for all financial-related threats, i.e., 
service slacking and theft, as well as the service corruption threat. For the 
service theft of file retrieval, which is the category that we asked the 
participants to investigate in Step 3, only three participants in the STRIDE 
session spotted this threat, while 17 subjects in the ABC session did so, or 
around 13\% and 71\%, respectively. Furthermore, none of STRIDE participants spotted 
the service theft of file storage and slacking of file retrieval, 
while only one participant spotted service slacking 
of file storage. On the other hand, 67\%, 17\%, and 50\% of ABC participants 
identified these categories, respectively. This, again, shows that the ABC threat 
classes guided the subjects toward service and payment related threats 
in a better way than the general categories included by STRIDE.

\subsubsection{Threat Model Accuracy}
To quantify accuracy, we compute the recall and precision values 
for the concrete threat scenarios found by  
each subject as compared to the reference threat models we built for 
ArchiveCoin. The recall is computed as 
$TP/(TP + FN)$, and precision is computed as $TP/(TP + FP)$, where 
a true positive $TP$ is a correctly 
identified threat, false negative $FN$ is an undetected 
threat, and false positive $FP$ is an incorrectly defined threat. The recall 
(precision) indicates how many valid (invalid) threats 
a subject defined. Both quantities take values between 0 and 1.

Based on the results of Step 3 in each framework (i.e., eliciting concrete 
threat scenarios), we found that participants who applied ABC produced a 
larger number of valid threat cases than the STRIDE session subjects, with  
average recall values of 0.48 and 0.4, respectively. At the same time, 
participants using ABC identified a lower number of irrelevant cases than 
those who applied STRIDE. The former scored an average 
precision value of 0.57, as opposed to 0.48 for the latter\footnote{One 
participant in the ABC 
session has 0 false negative and 0 true positive values, so we excluded him/her when 
we computed the average.}.

We believe that the result above can be attributed to several factors. 
First, ABC directed participants to consider the financial aspects of the 
system, which affected the elicited threat scenarios. Second, 
the use of the collusion matrices helped ABC participants reason about 
the threat space in an organized way that reduced random speculations, 
as opposed to the STRIDE threat tree patterns that work  
well when applied to software applications. Third, ABC's collusion matrices 
guided participants to spot threat cases caused by collusion, as opposed 
to STRIDE's tree patterns that focus only on solo attackers. The results show 
that none of the subjects in the STRIDE session identified a possible 
collusion case 
between a client and servers, while 11 subjects in the ABC session identified this 
collusion case\footnote{Most of them, however, did not provide 
a clear description of the attack scenario. Hence, these incomplete 
descriptions \emph{were not} counted as 
correct threats when grading Step 3.}. This confirms the importance of 
considering collusion when investigating threat cases, and shows 
the usefulness of ABC matrices in handling this task.

\begin{figure}[t!]
\centerline{
\includegraphics[height= 1.8in, width =0.35\columnwidth]{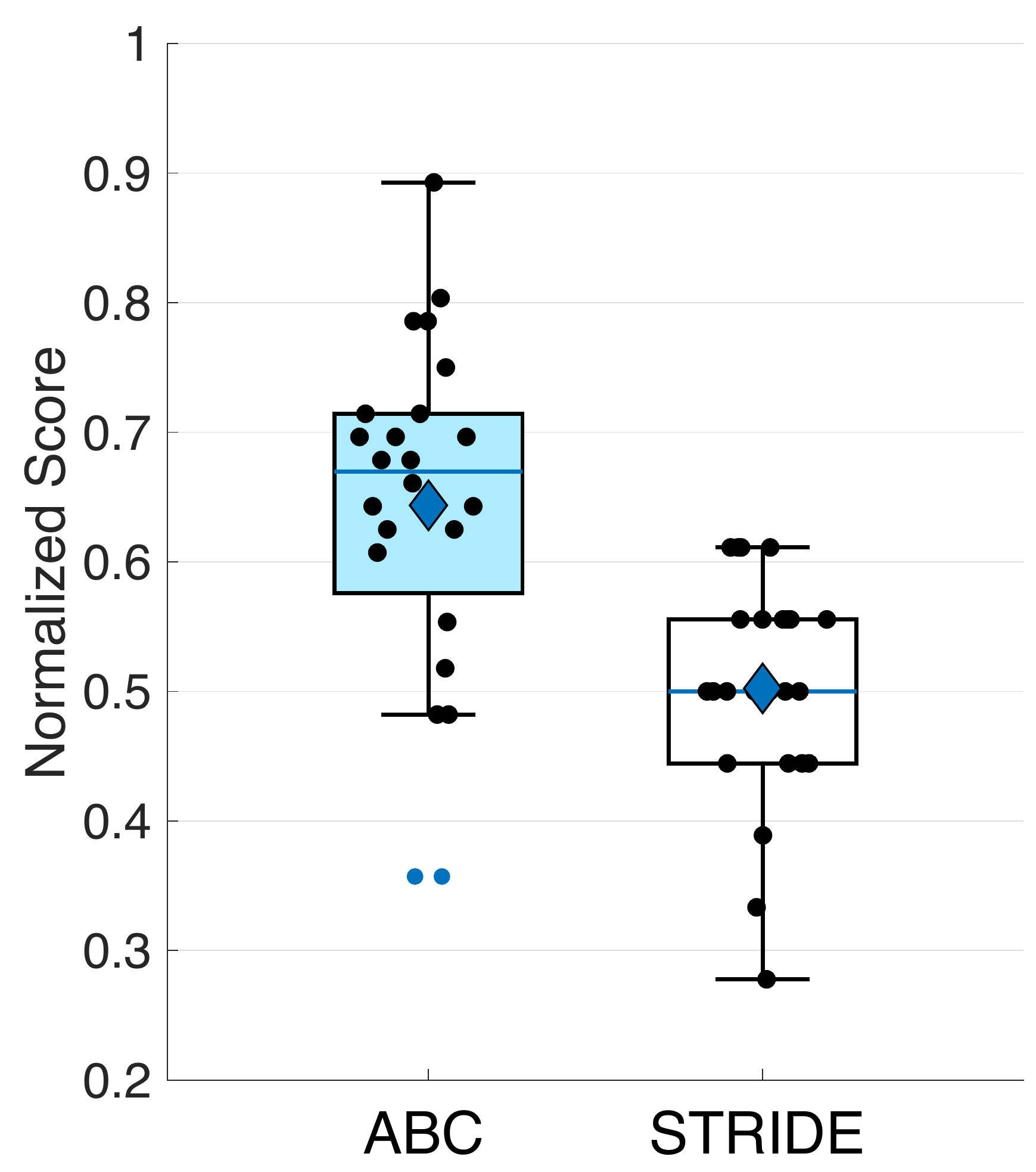}}
\caption{Total normalized scores (diamonds indicate the mean).}
\label{total-scores}
\end{figure}

All these factors affected the overall correctness of the threat models built 
by the subjects. As shown in Figure~\ref{total-scores}, the ABC session 
scored an average of 64$\%$ 
as compared to 50$\%$ for STRIDE. This is expected based on the reported  
results for the modeling steps, where ABC scores were ahead of those 
from STRIDE.

\subsubsection{Framework Ease-of-Use}
As mentioned previously, at the end of each session, we asked 
participants to report 
on how easy the framework in question was to apply. 
Ease-of-use was measured on a Likert scale in 
which 1 indicates the lowest value and 5 indicates the highest value. 
The average values are 3.9 for ABC and 3.8 for 
STRIDE\footnote{Three participants did not complete the 
questionnaire in the STRIDE session.}. This result becomes somewhat 
more significant when 
we point out that the participants already had some exposure to 
STRIDE and its threat tree patterns. Even 
though it is a new framework that introduced several new 
concepts to the participants, ABC still achieved a comparable ease 
of use level. This suggests that participants were able to grasp its 
concepts through just a single hour of training, and therefore 
ABC shows potential as a usable method for 
threat modeling.

\subsection{Threats to Validity} 
We acknowledge that a few limitations must be kept in mind when 
considering the study results. Empirical studies of threat 
modeling usually span a longer time frame, often on the orders of months, 
e.g., \cite{Scandariato15}. However, the fact that we were able to glean several 
important observations suggests that there may be lessons to 
be learned from short focused studies. In addition, 
we feel the design of our study might serve as a guide for determining 
promising areas for extended research before a large commitment of 
time and resources is made. Another 
constraining factor could be the age and experience level of our 
subjects. Different responses might have been 
obtained if we tested system security experts. However, we believe 
that the inexperience 
of our participants matches the cryptocurrency space well, which attracted 
users and researchers from various fields, even those from 
outside systems security. Therefore, our results give indications on how 
they might perform when investigating the security of 
cryptocurrencies.

\section{Experiences} 
\label{experiences} 
To demonstrate how ABC would function when applied to complex 
real-world systems, we developed use cases in which we 
built threat models for three cryptocurrencies, Bitcoin~\cite{bitcoin}, 
Filecoin~\cite{filecoin}, and our system CacheCash. Furthermore, 
we report on a non-cryptocurrency use case that  
targets a cloud native security technology called SPIFFE, a project worked 
on by one of our authors. Each of these 
cases represents a different stage in a system design lifetime.  
Bitcoin and SPIFFE are well-established systems, Filecoin is  
under development and close to being launched, and 
CacheCash is still in its early development stages. The analyses
for Bitcoin and Filecoin stopped before the risk management/mitigation 
phase. However, CacheCash and SPIFFE analyses involve the risk
mitigation step as we describe later.

\begin{table}[t!]
\caption{Threat model comparison.} 
\label{stats}
{\small
\begin{tabular}{| p{0.32\columnwidth}  | P{0.16\columnwidth} | P{0.16\columnwidth} | P{0.16\columnwidth}|P{0.16\columnwidth}|}\hline\hline
{\bf Aspect} & {\bf Bitcoin} & {\bf Filecoin} & {\bf CacheCash} & {\bf SPIFFE / SPIRE}   \\ [0.5ex] \hline\hline

ABC steps covered & Steps 1 - 3 & Steps 1 - 3 & Steps 1 - 4 & Steps 1 - 4  \\[0.5ex] \hline

Completion time (hr) & 10 & 47 & Not tracked & Not tracked \\[0.5ex] \hline

No. of collusion matrices & 5 & 14 & 9 & 4  \\ [0.5ex]  \hline   

Total threat cases & 105 & 882 & 525 & 1860  \\ [0.5ex]  \hline  

Distilled threat scenarios & 10 & 35 & 22 & 65  \\ [0.5ex]  \hline  

\end{tabular}}
\end{table}

\subsection{Bitcoin Analysis (Steps 1-3)} 
Bitcoin is by far the most valuable cryptocurrency with a capital market 
share of around $\$$245 billion as of  late June 2019~\cite{cap-market}. 
As shown in Table~\ref{stats}, the Bitcoin threat model has 
significantly fewer collusion matrices 
and threat cases than other systems\footnote{The full 
threat model for Bitcoin is available online \cite{material}}. This is because it  
provides only a currency exchange service, which reduces the number 
of assets. Furthermore, it involves only 
two types of participants, miners and clients, which reduces the size 
of the collusion matrices. These factors, in addition to our familiarity with 
Bitcoin design details, contributed in reducing the completion time of  
Bitcoin's threat model as shown in the table. Moreover, at the time we were 
working on this model, we had already completed the  
design of ABC. This suggests that deep 
understanding of the system model, and the availability of  
suitable tools impact not only the accuracy of the results, but 
also the time and effort expended in the threat 
modeling process.

We drew two main observations about the threat model we generated 
for Bitcoin. First, 
all the known threats to Bitcoin, such as double spending, Eclipse 
attacks~\cite{Heilman15}, Goldfinger attack~\cite{Kroll13}, and delaying 
blocks and transaction delivery~\cite{Gervais15}, were mapped to the 
collusion matrices produced by ABC.  
Second, collusion between participants can play a major role in Bitcoin's 
security. That is, several threats are neutralized by the assumption that 
at least 50$\%$ of the mining power is honest. Yet, mining pools have 
been formed to concentrate 
mining power. At the time of this writing, around 95$\%$ of the mining 
power is in the hands of just 10 mining pools~\cite{bitcoin-pools}. If   
the managers of these pools decide to collude, they could break the honest 
majority barrier and take the system down. In fact, serious security 
attacks can be performed with less amount of mining power. Sompolinsky et 
al.~\cite{Sompolinsky15} attest that a selfish mining 
attack, or blocks withholding, in which an attacker controls around 
30$\%$ of the mining power, would be able to undermine the fairness 
of the mining reward distribution.

Furthermore, miner collusion may take different forms, such as  
rejecting specific updates on the network protocol. Cryptocurrencies 
are still in the development stage, and so their protocols continue to 
receive new features and updates. These updates can create forks 
in the network~\cite{Antonopoulos17}, and should a subset of the 
miners agree not to adopt the modified protocol, a new version of the 
currency may be spun off. 
This happened in Bitcoin, where two new cryptocurrencies 
split from its network including Bitcoin Cash~\cite{bitcoincash} and Bitcoin 
Gold~\cite{bitcoingold}.

Usually, the $>$50\% threat, or miners' collusion in general, is argued 
about informally using incentive compatibility. This idea asserts that rational miners 
are more interested in keeping the system running to preserve 
the value of their rewards. However, this claim is hard to  
verify and remains as an open question~\cite{Bonneau15}. 
In addition, this assumption might be valid when all 
parties belong to the same system. Yet, miners could be working in 
several cryptocurrencies and it could be the case that destroying one  
to strengthen the other would be more profitable. Nonetheless, such 
observations highlight two key points. First, it indicates the 
importance of validating all the security assumptions a system makes in 
its design. And second, it points to the need for rational economic 
incentives to address some types of security threats that cannot 
be addressed by using only cryptographic approaches. 
The design of ABC accounts for such observations as mentioned 
earlier in this paper.

\subsection{Filecoin Analysis (Steps 1-3)} 
Filecoin~\cite{filecoin} is a cryptocurrency-based 
distributed file storage and retrieval system. Any party may join as a 
storage or retrieval miner to offer service to others. Filecoin operates 
distributed retrieval and storage markets where clients and 
miners can submit storage/retrieval bids and offers. Once these offers 
are matched, the service-payment 
exchange process, in which clients pay the miners in the Filecoin 
currency in exchange for receiving correct 
service, may start. The mining process in Filecoin  
is tied to the storage service miners put into the system. Recently, the Filecoin 
team raised around $\$$250 million   
through an ICO (initial coin offering)~\cite{filecoin-ico} in preparation 
for an official launch.

\begin{figure}[t!]
\centering
\subcaptionbox{Data storage service.\label{filecoin-net-model-storage}}{\includegraphics[width=0.45\columnwidth, height=1.4in]{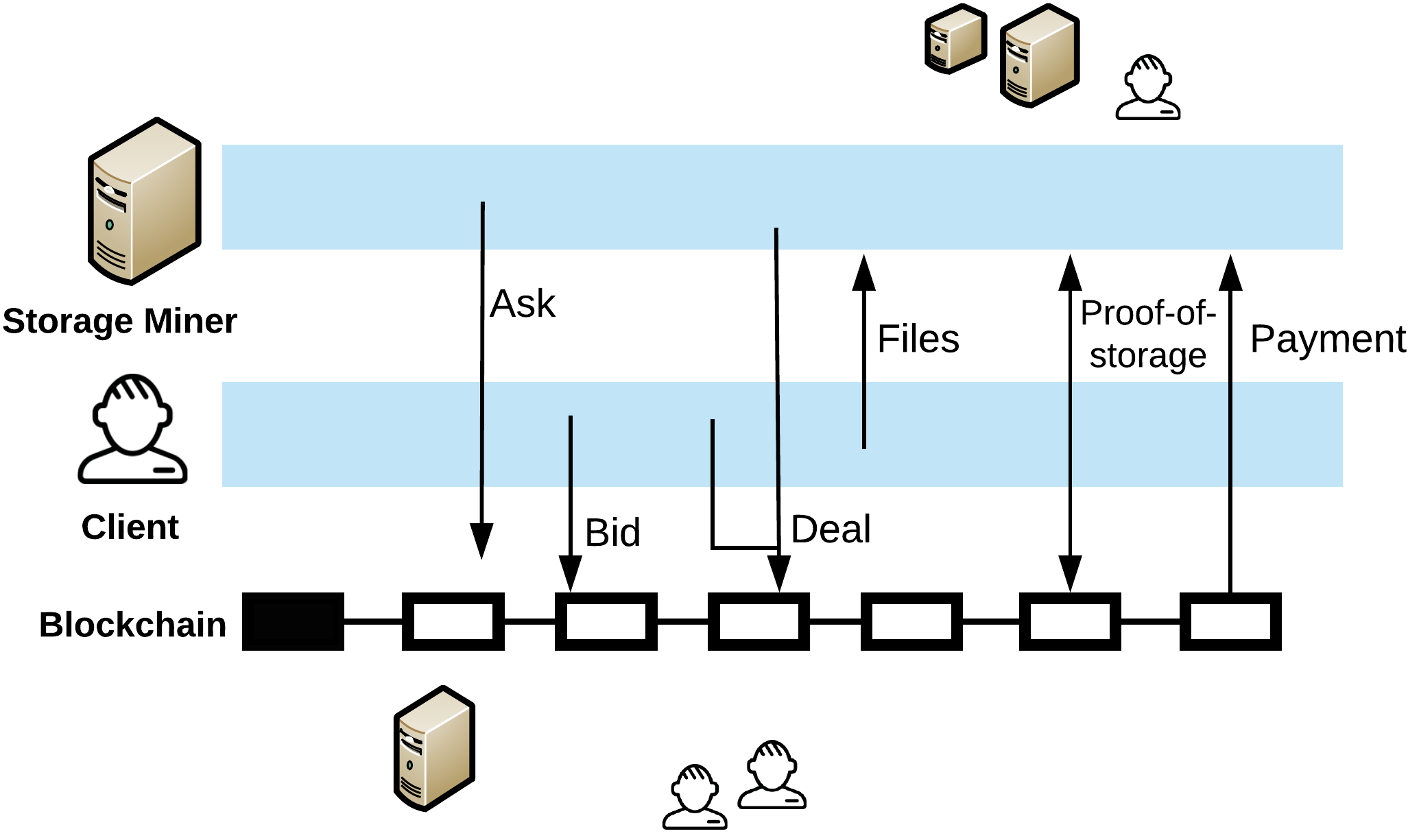}}\hfill
\subcaptionbox{Data retrieval service.\label{filecoin-net-model-retrieval}}{\includegraphics[width=0.45\columnwidth, height=1.4in]{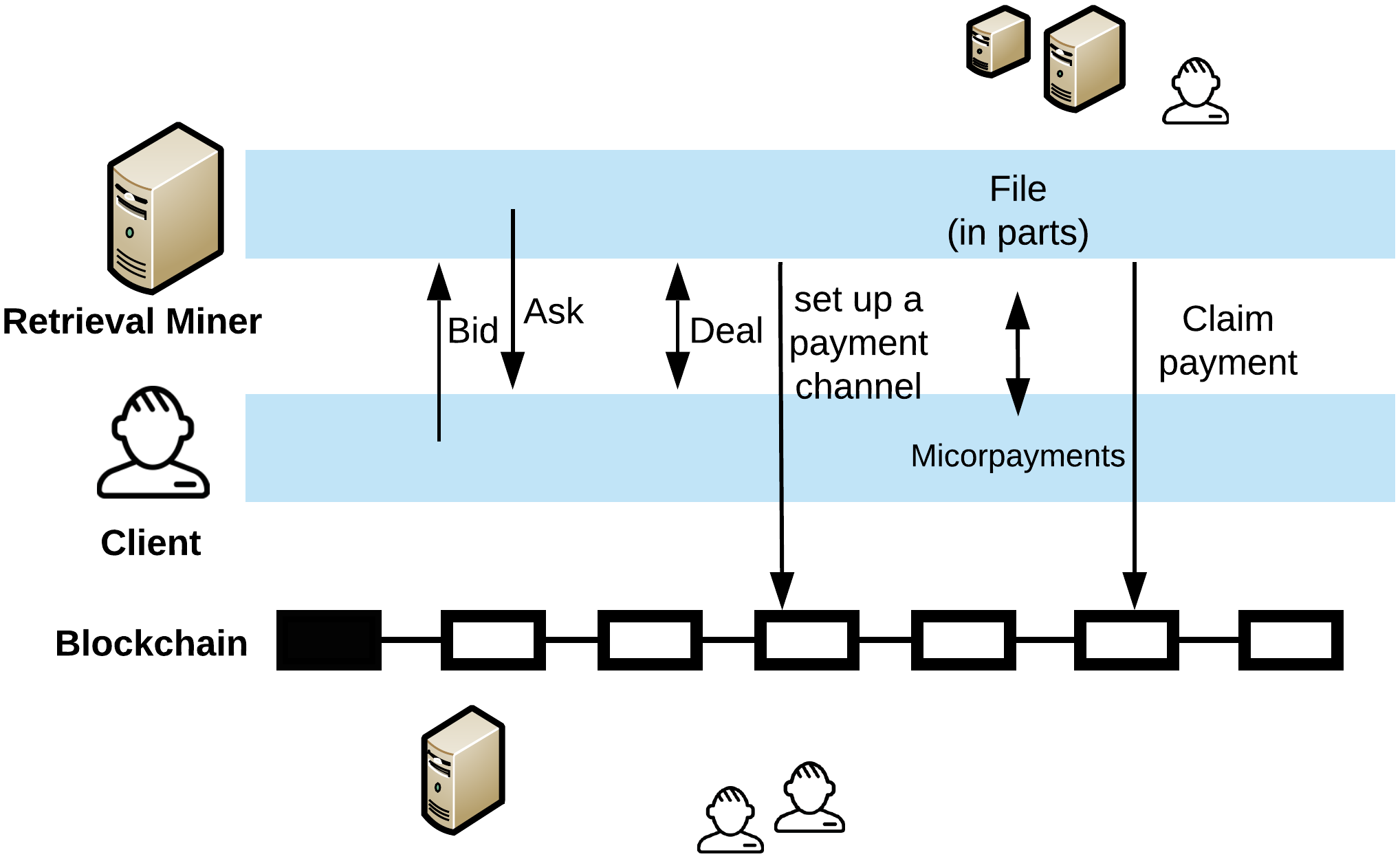}}\\

\caption{Filecoin network model.}
\label{filecoin-net-model}
\end{figure}

Filecoin is a more complicated system than Bitcoin as it provides 
two types of services on top of the currency exchange medium (these 
are captured by the network models shown in Figure~\ref{filecoin-net-model}). 
In addition, its protocol involves three participant roles: clients, retrieval miners, and 
storage miners, with the latter filling the traditional roles of miners in 
maintaining the blockchain. This complexity is reflected in 
the number of collusion matrices and threat cases produced as shown 
in Table~\ref{stats}. Moreover, all threat categories that target the service 
asset were replicated for each service type, which contributed to the large 
size of the threat model. These factors affected the completion time to build   
the model, which was 4.7x the time needed to build Bitcoin's model. This 
cost in time commitment is a natural result of working 
with newly developed and complex 
systems that provide a rich set of features.

In threat modeling Filecoin's whitepaper, we found 
three unaddressed issues, mostly dealing with collusion cases
that were not considered.  Additionally, there are many places where the
system is underspecified and so it is not possible to reason about whether
or not it meaningfully addresses a threat.\\

\noindent\emph{Ethics and disclosure.} 
We reached out to the Filecoin team, which 
mentioned efforts they have undertaken to resolve these problems. 
We withhold details about these issues
until later as part of the responsible disclosure process.

\subsection{CacheCash Analysis (Steps 1-4)}
\begin{figure}[t!]
\centerline{
\includegraphics[height= 2.0in, width =0.55\columnwidth]{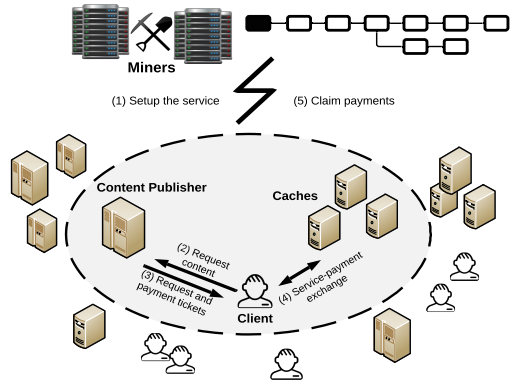}}
\caption{CacheCash network model.}
\label{cachecash-net-model}
\end{figure}

CacheCash is a cryptocurrency that provides a distributed content delivery 
service. It allows content publishers to construct dynamic 
networks of caches to serve their clients. This is done by allowing anyone 
to set up a new cache, and then collect cryptocurrency tokens from 
publishers for serving their clients (as captured by the network model shown 
in Figure~\ref{cachecash-net-model}). To handle the 
functionality of the currency exchange medium, CacheCash uses a 
modified version of the Bitcoin protocol.

We developed ABC during the early stages of our 
work on designing CacheCash. As the work progressed, we realized that 
most of the threat cases we encounter are related to 
the financial aspects of the system and the possible collusion  
between participants. Such aspects, as mentioned previously, are not 
explicitly addressed by  
traditional threat modeling frameworks. At that time we realized that 
none of these frameworks suited 
our needs, which lead to developing ABC.

As shown in Table~\ref{stats}, CacheCash's threat model is  
smaller, in terms of the number of collusion matrices and threat 
cases, than the one developed for Filecoin. This 
is because CacheCash provides a single type of service on top of 
the currency exchange medium, as compared to two in Filecoin. This 
is also reflected in the lower number of distilled threat cases than 
what Filecoin's model produced.

Beyond threat modeling, we used ABC while designing threat 
mitigation techniques in the CacheCash system. During that time, we observed the 
importance of rational financial incentives in this process. This includes 
employing detect-and-punish mechanisms  
in which the penalty deposit of a party is revoked upon 
detecting that it is cheating, or designing algorithms that, when implemented 
in a malicious way, can cost the attacker more in resources than 
would working honestly. Furthermore, we realized  
the value of game theory 
and economic analysis in assessing the effectiveness of these economic  
threat mitigation approaches, and in quantifying   
the risk, or amount of damage, that financial attacks may cause.  
To date, we found ABC useful for CacheCash in both the pre-design 
threat modeling step, and the after-design security analysis 
of the system modules.

As CacheCash's design is not public yet, its full threat model will be 
released at the public unveiling of the system in early 2020.

\subsection{SPIFFE/SPIRE Analysis (Steps 1-4)}
\begin{figure}[t!]
\centerline{
\includegraphics[height= 2.0in, width =.65\columnwidth]{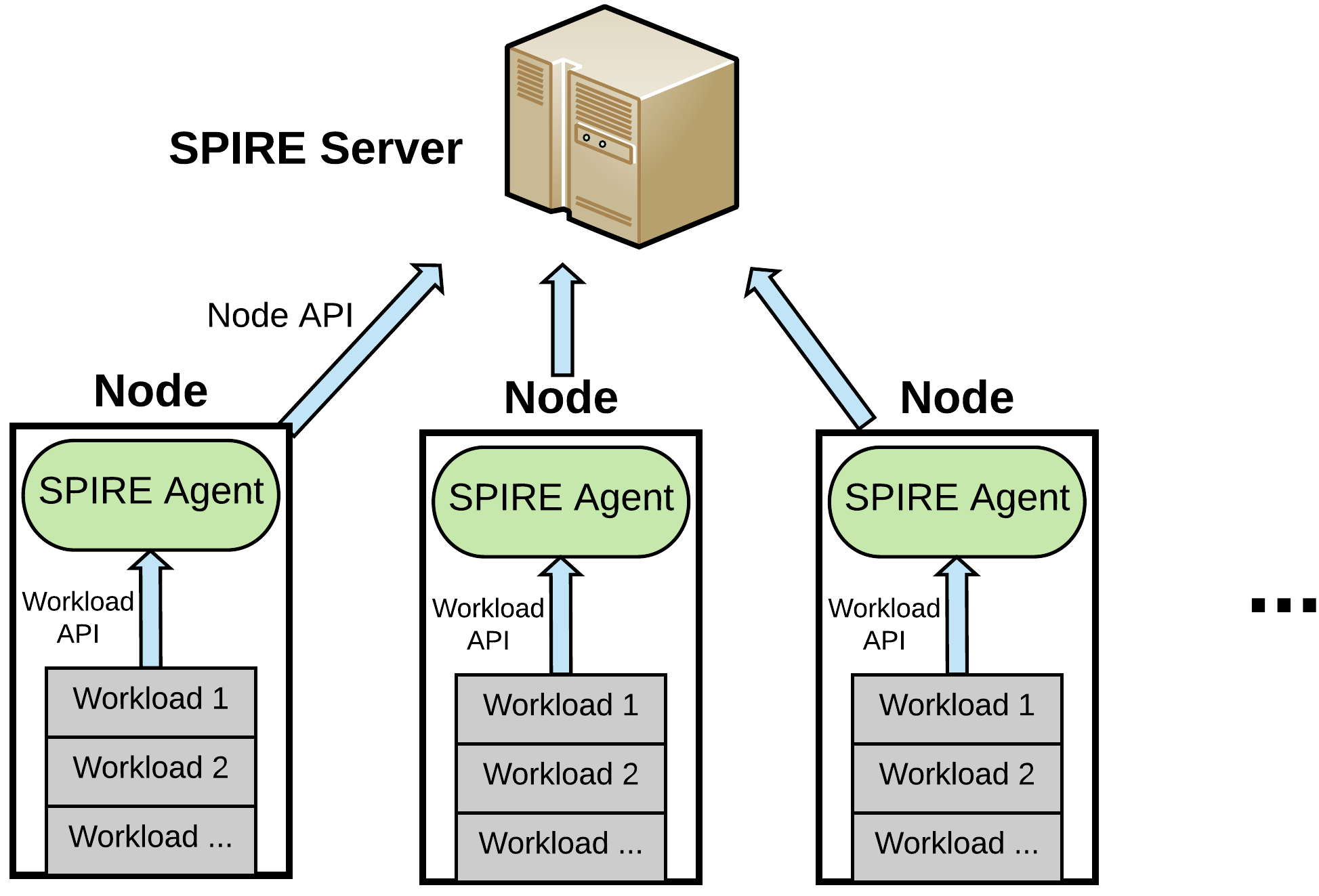}}
\caption{SPIFFE/SPIRE network model.}
\label{spiffe-net-model}
\end{figure}

The Secure Production Identity Framework for Everyone (SPIFFE) is a 
technology that targets the problem of obtaining identities in cloud-based 
distributed systems. SPIFFE, which is a project under the umbrella 
of the Linux Foundation, allows a service to 
acquire   
a secure identity that can be used to authenticate itself and 
authorize its access to other services and system resources. 
Thus, it solves the scalability problem of conventional security practices 
when working across 
heterogeneous environments and organizational boundaries. The SPIFFE 
Runtime Environment (SPIRE) is the production-ready implementation of 
SPIFFE APIs. Its work model, depicted in Figure~\ref{spiffe-net-model}, 
consists of a central server that mints identity documents for 
applications (or workloads), and node agents that perform workload attestation and 
distribution of IDs. 

The SPIFFE/SPIRE team collaborated with one of the authors of this paper in using 
ABC to evaluate the framework's 
security properties and to understand the potential risks in 
various practical deployment scenarios. The complete process can be found 
in~\cite{spiffe-part1, spiffe-part2}. This use case is different from the previously 
discussed ones because it does not involve any form of explicit payment exchange 
between the participants. Thus, it attests to the applicability of ABC for 
not only cryptocurrencies, but also for large-scale distributed systems in general.

Through their analysis, the SPIFFE team members produced four threat categories:  
misrepresentation of identity, identity theft, compromise/remote code execution, 
and DoS. They also outlined the specific capabilities, or strength classes, an 
attacker may need to perform a given attack. Such an attacker could be the SPIRE 
server, a node agent, a workload (on 
the same node or on a different node), or even 
an external entity. Listing the attacker strength classes helped in both enumerating the 
possible threat scenarios, as well as in estimating the likelihood and severity of 
each distilled threat case.

As shown in Table~\ref{stats}, although the threat model of SPIFFE has the least 
number of collusion matrices, it produced the largest number of threat cases (around 17.7x, 2x, 3.5x 
of what Bitcoin, Filecoin, and CacheCash produced, respectively). The lower number of 
matrices is due to the fact that all 
other systems have the currency exchange medium, which adds another layer beneath the 
distributed service they provide. SPIFFE, on the other hand, provides only an identity production 
service that can be run as an additional layer on top of any cloud native application. However, 
SPIFFE has 4 participant roles (on the attacker side it would be 5 when counting the 
external entity). This made its collusion matrices larger, and hence, resulted in more 
threat cases than the other systems.

Similar to the CacheCash use case, ABC was used during the risk management and threat 
mitigation step. The team members computed a score for each of the distilled threat scenarios that 
estimates the likelihood that an attacker possesses the required capabilities, and the 
severity of the attack to the system. Studying the list of scored threat scenarios,  
the model outlined 54 unaddressed threat cases, most of which have very low scores, and produced 
several insights that guided the team on where to focus their efforts. Prior to the 
threat modeling process, the team envisioned several threat mitigation 
mechanisms that could complicate the system design. Once they 
determined that the impact of such attacks is very low, the higher score attacks 
were considered instead. These scenarios required even simpler mitigation 
techniques. For example, one type of DoS attacks can be mitigated using a 
rate limit control mechanism that 
the team integrated with the system~\cite{spire-rate-limiter}.

Accordingly, this case shows how ABC can be very useful in assessing the security 
level of an already deployed system, defining which threat cases are already neutralized, and 
mitigating high profile, previously unaddressed, attacks.

\section{Conclusions}
\label{conclusions}
In this paper, we introduce ABC, a cryptocurrency-focused threat modeling 
framework. Its design is motivated by the observation that 
traditional threat modeling frameworks do not fit  
cryptocurrencies, thus leaving them vulnerable to unanticipated attacks. ABC 
introduces collusion matrices, a technique that allows designers to investigate
hundreds of threat cases in a reasonable amount of time. This is in addition to a 
flexible mechanism to derive system specific threat categories that focus on  
the assets to be protected and account for the financial motivations of the attackers. 
Both the user study and the use cases confirm that our 
framework is effective in unraveling hidden threat cases. The 
evaluated cases cover various types of cryptocurrency-based systems and 
a cloud native security technology under the Linux Foundation. This shows the 
potential of ABC to improve the security of a wide array of distributed systems.

{\footnotesize \bibliographystyle{acm}
\bibliography{abcBib}}

\begin{appendices}
\renewcommand{\thesection}{\Alph{section}}
%the following adds the word Appendix before the letter numbering
%\renewcommand{\thesection}{\appendixname~\Alph{section}}

\section{Deriving ABC Threat Categories}
\label{threat-categories}
In this section, we show how the threat categories listed in Table~\ref{threat-cat} 
were derived. We apply the procedure outlined in Step 2 of the ABC framework 
(Section \ref{step2}) to the assets of CompuCoin, which include the service (e.g., 
computation outsourcing in case of CompuCoin), service 
rewards or payments, blockchain, transactions, 
currency, and communication network. For each asset, we outline its security 
properties in order to identify any factors that might violate these properties. 
These factors are labeled as threat categories.

Starting with the {\bf service asset}, our analysis outlines the following:
\begin{enumerate}
\itemsep0em
\item \emph{Security properties}: A secure service can be defined as 
the action of  
serving clients correctly at anytime, while providing confidentiality 
and binding the servers to the service they provide. Hence, 
the security properties of a service asset may include integrity, 
availability, confidentiality, and non-repudiation.

\item \emph{Threat categories:} By negating the above properties, 
we find that the service asset has the following threat categories:
\begin{itemize}
\itemsep0em
\item Service tampering/corruption: An attacker provides clients with 
invalid service or corrupts the correct service delivered by others.

\item Information disclosure: An attacker reveals the contents of 
service-related messages, such as the service content/outcome, the 
service requests sent by clients, replies sent by servers, etc. 

\item Repudiation: A server denies providing a specific service or 
a client denies receiving it.

\item DoS: An attacker makes the service unavailable to legitimate 
users.
\end{itemize}
\end{enumerate}

Next, we analyze the {\bf service payment} (or rewards) asset as follows:
\begin{enumerate}
\itemsep0em
\item \emph{Security properties:} One may consider the service 
payment asset secure as long as: a) servers are rewarded properly 
for the service they provide, and b) servers earn the payments 
they collect.

\item \emph{Threat categories:} Negating the above security 
requirements produce the following threat categories:
\begin{itemize}
\itemsep0em
\item Service slacking: A server collects payments without 
performing all the promised work.
\item Service theft: A client obtains service for a lower payment 
than the agreed upon amount. 
\end{itemize}
\end{enumerate}

For the {\bf blockchain asset}, our analysis produces the following:
\begin{enumerate}
\itemsep0em
\item \emph{Security properties:} The blockchain security properties 
are tied to the security of the underlying consensus protocol. These properties 
have been thoroughly studied in the literature \cite{Bonneau15, Garay15, Pass17}. 
We adopt the ones introduced in \cite{Pass17} with slight modifications 
based on the work presented in \cite{Bonneau15}, which include:
\begin{itemize}
\itemsep0em
\item Consistency: At any point in time, honest miners hold copies of the 
blockchain that have a common prefix and may differ only in the last $y$ 
blocks, where $y$ is a block confirmation parameter. A block then is 
confirmed once it is buried under $y$ blocks on the blockchain.

\item Future-self consistency: At any two points in time, $t_1$ and $t_2$, the 
blockchain maintained by an honest party may differ only in the last 
$y$ blocks. Consistency and future-self consistency properties achieve 
blockchain persistence or immutability. 

\item Fairness: Miners collect mining rewards in proportion to the 
resources they expend in the mining process.

\item Correctness: All the blocks within the longest branch in the 
blockchain are valid. (Note that correctness and fairness represent 
the chain quality property outlined in \cite{Garay15, Pass17}.) 

\item Growth: As long as the system is functional, new valid blocks 
will be added to the blockchain.
\end{itemize}

\item \emph{Threat Categories:} By negating the above properties, we 
can distill the following threat categories for the blockchain asset:
\begin{itemize}
\itemsep0em
\item Inconsistency: Honest miners do not agree on the prefix of the 
blockchain copies they hold beyond the unconfirmed blocks. This 
also covers the case of an honest miner who does not agree with 
itself on the blockchain prefix it holds over time, e.g., alternating 
between two branches that compete in being the longest.

\item Invalid block adoption: The longest chain contains 
corrupted blocks that either have an invalid format or contain 
invalid transactions.

\item Biased mining: A miner pretends to expend the needed resources 
to be selected to extend the blockchain and collect the 
mining rewards.

\item Chain freezing: The blockchain does not grow at a regular rate, 
but instead freezes for several contiguous rounds. 
This threat category is a form of DoS attack, and hence, we cover it under 
DoS against the communication network asset.
\end{itemize}
\end{enumerate}

Next we analyze the {\bf transaction asset} as follows:
\begin{enumerate}
\itemsep0em
\item \emph{Security properties:} Secure transactions can be characterized as 
correct, tamper-proof, and source-binding, i.e., cannot be denied by the 
originator. In addition, these transactions need to be accessible to 
the system users at any time so they can send/receive/view transactions as  
needed. Moreover, these transactions must not 
reveal any information about the source, destination, and amount of 
transferred funds. Accordingly, we outline the following security 
properties for the transaction asset: non-repudiation, integrity, validity, 
availability, and anonymity. Note that the validity property is already covered by the 
correctness aspect of the blockchain, where a valid blockchain contains 
only valid transactions. Furthermore, the availability property is covered 
under the communication network asset.

\item \emph{Threat categories:} Based on the previous discussion, and 
again by negating the aforementioned security properties, the 
threat categories for the transaction asset would be:
\begin{itemize}
\itemsep0em
\item Repudiation: An attacker denies issuing transactions.

\item Tampering: An attacker manipulates the fields of a transaction.

\item Deanonymization: An attacker violates users' privacy by 
exploiting the public nature of the blockchain to link transactions and 
payments, and use this knowledge to track   
the activity of these users in the system and, possibly, reveal their real 
identities.
\end{itemize}
\end{enumerate}

Next we analyze the {\bf currency asset} as follows:
\begin{enumerate}
\itemsep0em
\item \emph{Security properties:} The security properties of the 
currency asset are intertwined with the properties of the 
transaction asset. This is because the currency takes the form of 
digital tokens, which are the transactions exchanged in the 
system. Thus, they inherit all the transaction security properties. 
What remains is to deal with the currency ownership, 
meaning that only the owner can spend these tokens. 

\item Threat categories: Beside the categories outlined above 
for the transaction asset, we have the following threat category 
for the currency asset:
\begin{itemize}
\itemsep0em
\item Currency theft: An attacker steals currency from others in 
the system. This includes all currency theft attacks that 
are not covered by other assets. For example, biased mining, 
where a miner steals others rewards indirectly,  is  
currency theft, 
but it is already covered by the blockchain. The same holds true  
for the service payment related threats. 
\end{itemize}
\end{enumerate}

Finally, we analyze the {\bf communication network asset} as follows:
\begin{enumerate}
\itemsep0em
\item \emph{Security properties:} The communication network is the 
backbone of any cryptocurrency system, and one that is unreliable 
can lead to numerous problems.   
First, it can create delays in propagating newly mined blocks that could produce 
an inconsistent blockchain. Second, it can cause delays in relaying 
transactions, which could reduce the transaction throughput of the 
system and affect its availability aspect. Third, it can slow down setting up new 
miners who need a longer time to discover other peers and download 
copies of the blockchain. Fourth, it opens the possibility of being controlled by 
external parties that could intercept the communication links and 
isolate nodes in the network. Consequently, a secure cryptocurrency 
system needs a reliable and robust communication network. We 
merge all these aspects into one security property, namely, availability. 

\item \emph{Threat categories:} The communication network asset has 
one threat category, which is DoS.
\end{enumerate}

Table~\ref{threat-cat} summarizes all the threat categories derived in 
this appendix. As mentioned previously, this table is by no means 
comprehensive. Additional threats can be added based on the asset 
types of the system, or 
more refined definitions of the asset security properties. 
This detailed treatment was provided as 
a thorough example to clarify the application of Step 2 in the ABC 
framework. Nonetheless, we found this table sufficiently detailed when 
building threat models 
for the systems reported as use cases in Section~\ref{experiences}, 
including Bitcoin, Filecoin, and CacheCash, as well as for the user study 
tutorial as reported in Section~\ref{evaluation}. For this reason, 
Table~\ref{threat-cat} can be viewed as a base threat list that can be 
extended, or even reduced, based on the system under design.   

\end{appendices}

%\theendnotes

\end{document}